\documentclass[10pt, twocolumn, nofootinbib,superscriptaddress]{revtex4-1}
\usepackage{amsmath,amssymb,amsfonts}
\usepackage{algorithmic}
\usepackage{graphicx}
\usepackage{textcomp}
\usepackage{xcolor}
\usepackage{ragged2e}
\usepackage{booktabs, makecell, tabularx}

\usepackage{gensymb}

\makeatletter
    \renewcommand\@make@capt@title[2]{%
     \@ifx@empty\float@link{\@firstofone}{\expandafter\href\expandafter{\float@link}}%
      {\textbf{#1}}\@caption@fignum@sep#2\quad}%
\makeatother
\makeatletter 
\renewcommand{\fnum@figure}{\textbf{Figure~\thefigure}}
\makeatother

\usepackage{xcolor}
\newcommand{\beginsupplement}{%
        \setcounter{table}{0}
        \renewcommand{\thetable}{S\arabic{table}}%
        \setcounter{figure}{0}
        \renewcommand{\thefigure}{S\arabic{figure}}%
     }
\def\BibTeX{{\rm B\kern-.05em{\sc i\kern-.025em b}\kern-.08em
    T\kern-.1667em\lower.7ex\hbox{E}\kern-.125emX}}

\begin{document}

\title{Guided-Acoustic Stimulated Brillouin Scattering in Silicon Nitride Photonic Circuits}

\author{Roel~Botter}
\author{Kaixuan~Ye}
\author{Yvan~Klaver}
\author{Radius~Suryadharma}
\author{Okky~Daulay}
\author{Gaojian~Liu}
\author{Jasper~van~den~Hoogen}
\author{Lou~Kanger}
\author{Peter~van~der~Slot}
\affiliation{Nonlinear Nanophotonics, MESA+ Institute of Nanotechnology, University of Twente, Enschede, the Netherlands}
\author{Edwin~Klein}
\author{Marcel~Hoekman}
\author{Chris~Roeloffzen}
\affiliation{LioniX International, Enschede, the Netherlands}
\author{Yang~Liu}
\affiliation{Institute of Physics, Swiss Federal Institute of Technology Lausanne (EPFL), Lausanne, Switzerland}
\author{David~Marpaung}
\email{Corresponding author: david.marpaung@utwente.nl}
\affiliation{Nonlinear Nanophotonics, MESA+ Institute of Nanotechnology, University of Twente, Enschede, the Netherlands}

\date{\today}

\begin{abstract}
Coherent optomechanical interaction between acoustic and optical waves known as stimulated Brillouin scattering (SBS) can enable ultra-high resolution signal processing and narrow linewidth lasers important for next generation wireless communications, precision sensing, and quantum information processing. While SBS has recently been studied extensively in integrated waveguides, many implementations rely on complicated fabrication schemes, using suspended waveguides, or non-standard materials such as As$_2$S$_3$. The absence of SBS in standard and mature fabrication platforms prevents large-scale circuit integration and severely limits the potential of this technology. Notably, SBS in standard silicon nitride integration platform is currently rendered out of reach  due to the lack of acoustic guiding and the material's infinitesimal photo-elastic response. In this paper, we experimentally demonstrate advanced control of backward SBS in multilayer silicon nitride waveguides. By optimizing the separation between two silicon nitride layers, we unlock gigahertz acoustic waveguiding in this platform for the first time, leading up to 12-15~× higher SBS gain coefficient than previously possible in silicon nitride waveguides. Using the same principle, we experimentally demonstrate on-demand inhibition of SBS by preventing acoustic guiding in the waveguide platform. We utilize the enhanced SBS gain to demonstrate a microwave photonic notch filter with high rejection (30 dB). We accomplish this in a low-loss, standard, and versatile silicon nitride integration platform without the need of suspending the SBS-active waveguide or hybrid integration with other materials. 
\end{abstract}

\maketitle

\section*{Introduction}

\begin{figure*}[t!]
\centerline{\includegraphics[width=0.9\textwidth]{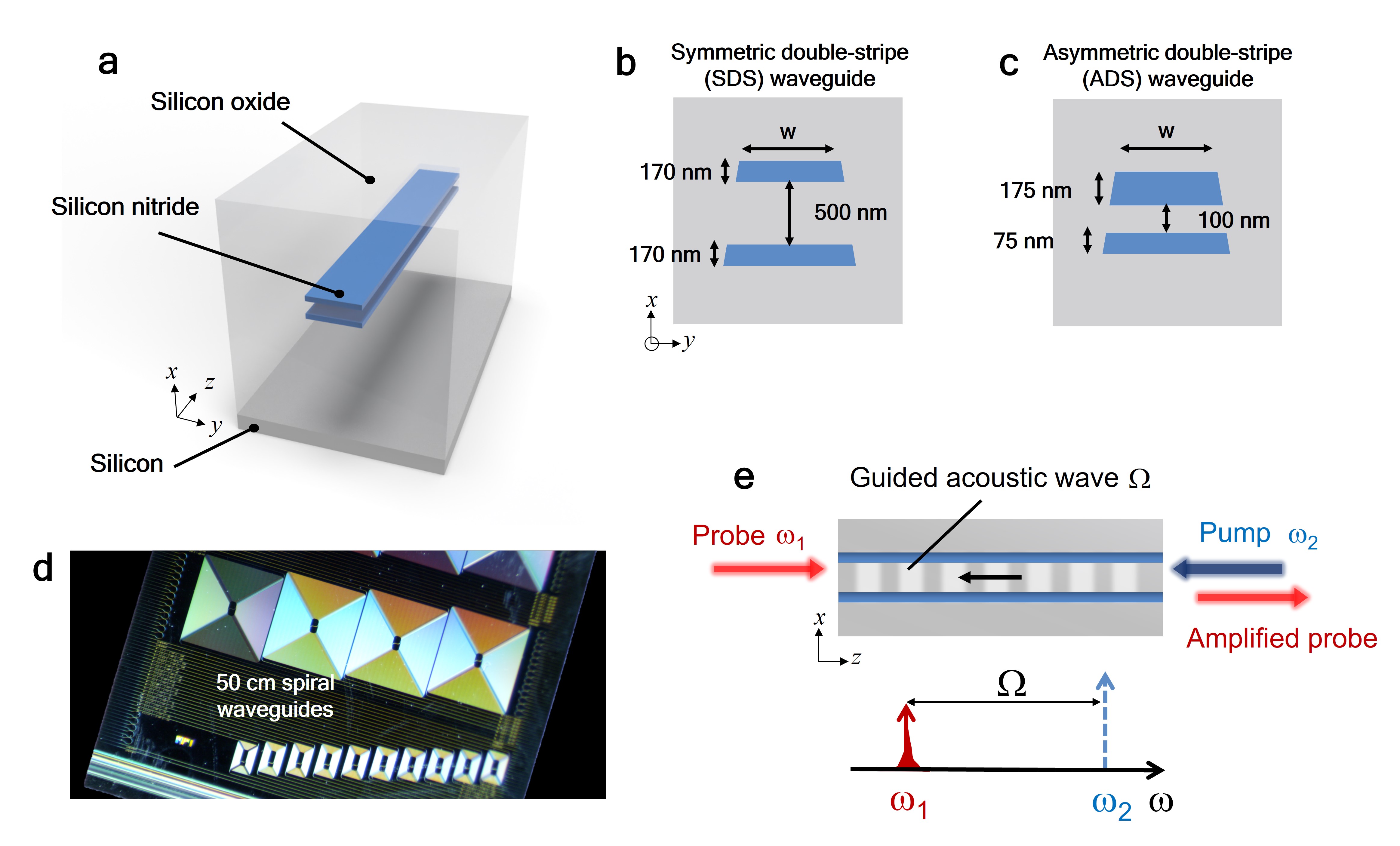}}
\caption{\textbf{Multilayer silicon nitride waveguides}. \textbf{(a)} Artistic representation of the waveguide consisting of two layers of silicon nitride separated by a layer of silicon oxide. \textbf{(b-c)} Diagrams of the symmetric double stripe (SDS) and assymetric double-stripe (ADS) waveguide cross-section and the dimension. Blue indicates the silicon nitride core, and grey indicates the silicon oxide cladding. \textbf{(d)} Photograph of the silicon nitride sample used in the experiments. Each of the large squares contains two waveguide spirals, each 50 cm long. Each square contains a waveguide of a different width. \textbf{(e)} Diagram illustrating backward stimulated Brillouin scattering (SBS) in the SDS silicon nitride waveguide. The optical pump at $\omega_{2}$ is converted to the Stokes-shifted probe light at $\omega_{1}$ by the gigahertz acoustic waves, with a frequency $\Omega=\omega_{2}-\omega_{1}$, guided between the two silicon nitride layers.}
\label{fig:fig1}
\end{figure*}

Brillouin optomechanics is on the rise; coherent control of light through optomechanical interactions  \cite{Aspelmeyer2014CavityOptomechanics,Safavi2019,Eggleton2019BrillouinPhotonics} with acoustic waves and vibrations is a burgeoning field with a wide range of applications, from quantum optics to telecommunications. In particular, Stimulated Brillouin scattering (SBS)  \cite{Boyd2008NonlinearOptics,Rakich2012GiantLimit}, which is an optomechanical interaction between light and gigahertz sound waves, is currently revolutionizing integrated optics  \cite{Eggleton2019BrillouinPhotonics}. SBS spectrally manifest in narrowband (tens of MHz) gain resonance, shifted in frequency by about 10 GHz from the pump, making it a unique filter and amplifier with technological importance in next generation optical and radio communications \cite{Marpaung2019IntegratedPhotonics,Marpaung2015Low-powerSelectivity,Li2013MicrowaveOscillator,Gertler2020TunableSilicon,Otterstrom2019ResonantlyAmplifier}, high precision sensors \cite{Li2017MicroresonatorGyroscope,Lai2020EarthGyroscope,Bashan2018OptomechanicalReflectometry,Denisov2016GoingDemonstration}, laser sources with ultra-narrow linewidth \cite{Gundavarapu2018Sub-hertzLaser,Otterstrom2018ALaser,Kabakova2013NarrowChip,Al-Taiy2014Ultra-narrowSpectroscopy}, and non-reciprocal control of light propagation \cite{Kittlaus2020ElectricallyPhotonics,Kittlaus2018Non-reciprocalModulation,Sohn2018Time-reversalCircuits,Kim2015Non-reciprocalTransparency}.

Since it's discovery, SBS has been extensively studied in optical fibers \cite{Thevenaz2008SlowFibres,Okawachi2005TunableFiber,Zhu2007StoredScattering,Ippen2003StimulatedFibers}, benefiting from their long length and low loss. But the last 10 years have brought a real paradigm shift showing that SBS can be induced and enhanced in centimeter-length chip-scale devices \cite{Eggleton2019BrillouinPhotonics,Pant2011On-chipScattering,VanLaer2015InteractionNanowire,Shin2015ControlEmitter-receivers,Kittlaus2016LargeSilicon,Kittlaus2017On-chipScattering,Sohn2018Time-reversalCircuits}. Recent demonstrations have produced SBS devices in chalcogenide glasses, silicon, silica, gallium arsenide, and aluminum nitride. This progress entails a tantalizing promise with profound impacts: the integration of SBS devices in large-scale and versatile circuits for various classical and quantum information processing technologies. But despite encouraging results, SBS are devices still singular and difficult to interconnect due to the need of unconventional materials \cite{Pant2011On-chipScattering} or suspending the devices in air to avoid leakage of acoustic waves \cite{Kittlaus2016LargeSilicon}. 

Silicon nitride has emerged as a leading photonic integrated circuit (PIC) platform due to its versatility \cite{Zhuang2011Low-lossProcessing,Roeloffzen2018Low-LossOverview,Roeloffzen2013SiliconCircuits} and exceptional performance in terms of linear and nonlinear losses \cite{Puckett2021422Linewidth,Spencer2014IntegratedRegime,Kordts2016PhotonicPhotonics,Worhoff2015TriPleX:Platform}. It has been the material of choice for recent important demonstrations including narrow linewidth lasers \cite{Gundavarapu2018Sub-hertzLaser,Xiang2019Ultra-narrowGrating,Fan2020HybridLinewidth,Jin2021Hertz-linewidthMicroresonators} and optical frequency combs \cite{Shen2020IntegratedMicrocombs,Gaeta2019Photonic-chip-basedCombs,Stern2018Battery-operatedGenerator,Moss2013NewOptics}. The low loss and compact silicon nitride circuits also underpin large scale signal processing for microwave photonics \cite{Marpaung2019IntegratedPhotonics,Roeloffzen2013SiliconCircuits,Marpaung2013IntegratedPhotonics} and integrated  quantum photonics \cite{Taballione20198x8Waveguides, Mehta2020IntegratedLogic,Arrazola2021QuantumChip,Niffenegger2020IntegratedQubit} and drive the surging research in programmable photonics and photonics artificial intelligence \cite{Zhuang2015ProgrammableApplications,Rios2015IntegratedMemory,Shastri2021PhotonicsComputing,Feldmann2021ParallelCore}. The wide transparency window of silicon nitride also enable applications in the visible wavelength regime with immediate applications in biological imaging \cite{Munoz2019FoundryMid-Infrared, Romero-Garcia2013SiliconWavelengths, Moreaux2020IntegratedTime} and in technologies based on trapped atomic ions for high fidelity quantum information processing and high-accuracy portable atomic clocks \cite{Mehta2020IntegratedLogic, Mehta2016IntegratedQubit, Niffenegger2020IntegratedQubit, West2019Low-lossRegime}. 

Recently, there is increasing interest towards SBS actuation in silicon nitride waveguides. To date, there have been two pioneering reports of SBS gain measurements in standard silicon nitride waveguides, where the silicon nitride core is not released from the cladding and substrate materials. SBS measured in an thin (40 nm thickness) silicon nitride waveguide with a very large mode area (also referred as dilute nitride waveguide) yield a gain coefficient of 0.1 m$^{-1}$W$^{-1}$ \cite{Gundavarapu2018Sub-hertzLaser}. On the other hand, SBS in thick (800 nm) silicon nitride waveguide  yield an even lower gain coefficient of 0.07 m$^{-1}$W$^{-1}$ \cite{Gyger2020ObservationWaveguides}, which is below the gain level achieved in optical fibers. Both demonstrations are plagued by acoustic leakage from the silicon nitride core to the surrounding silicon oxide, preventing the benefit of SBS enhancement in nanophotonic waveguides \cite{Eggleton2019BrillouinPhotonics}. Such a low gain severely limits the application of SBS in this waveguide platform only to  SBS lasers.  The absence of SBS in standard silicon nitride waveguides leaves a massive technological gap in an already mature and rich technological platform. Strong Brillouin optomechanics in silicon nitride will lead to signal processing, sensing, and light source technologies relevant for next generation wireless communications, computing, and quantum information processing. 

In this work we demonstrate, for the first time, guided-acoustic wave SBS in multi-layer silicon nitride nanophotonic circuits. We use multilayer silicon nitride waveguides to confine both the optical and the gigahertz acoustic waves, leading to enhanced SBS gain and narrower resonance linewidth, unlike in previous demonstrations. Through simulations and experiments, we show the feasibility of tailoring the cross-section of the multilayer silicon nitride waveguides for on-demand enhancement or inhibition of SBS, opening the pathway towards circuit-level selective SBS generation. We also demonstrate the first application of SBS in silicon nitride for RF signal processing through a high-extinction (30 dB) and high spectral resolution (400 MHz) notch filter using only 0.4 dB of SBS gain. Finally we present the pathways to record-high SBS gain in standard silicon nitride waveguides without the need of suspending the devices in air. Our results open the way to integrating unique SBS element in a large-scale circuit and intersecting it with emerging technologies including narrow linewdith lasers, frequency combs, and programmable photonic circuits.  

\section*{Results}

\subsection*{Multilayer Silicon Nitride Waveguides}\label{sec:results}
An artistic representation of the  multilayer silicon nitride waveguide explored in this work is depicted in Fig.~\ref{fig:fig1}~\textbf{(a)}. This structure consists of two silicon nitride layers separated by a silicon oxide layer which is also the material of the upper and bottom cladding (See Methods for details of the fabrication). The waveguide geometry has been optimized for simultaneously achieving tight bend radius (of the order of 100 µm) and low propagation loss (of the order of 0.1 dB/cm) \cite{Roeloffzen2018Low-LossOverview}. For this reason, such a  multilayer waveguide has been widely used in large-scale high complexity circuits for microwave photonics \cite{Marpaung2019IntegratedPhotonics}, integrated quantum photonics, and integrated narrow linewidth lasers. 

Two distinct waveguide geometries have been investigated here: first, the so-called symmetric double stripe (SDS) waveguide that consists of two 170 nm thick layers of silicon nitride, separated by a 500 nm layer of silicon oxide, as shown in Fig.~\ref{fig:fig1}~\textbf{(b)}. For a 1.2~µm-wide waveguide at a wavelength of 1550~nm, this results in a mode field diameter of 1.6 × 1.7 µm (x × y) and an effective index of 1.535. The waveguide can be coupled to a single-mode fiber by tapering the silicon nitride height to 35~µm.The second waveguide geometry explored here asymmetric double stripe (ADS) design as shown in Fig.~\ref{fig:fig1}~\textbf{(c)}. The silicon nitride layers in this geometry are of different heights, 75 nm for the bottom layer, and 175 nm for the top layer. The intermediate silicon oxide layer has been reduced to 100 nm. This waveguide has a mode field diameter of 1.5 × 1.2 µm (x × y), and the effective index is 1.535, just as for the SDS waveguide. 

The samples are 50 cm long spirals of both SDS and ADS waveguides, Fig \ref{fig:fig1}~\textbf{(d)} shows a photograph of the SDS sample. The spirals are tightly wounded with 100~µm bend radius, showing the feasibility of dense circuit integration in this high waveguide platform. This is in stark contrasts with dilute-mode thin silicon nitride platform with bend radius of a few mm \cite{Gundavarapu2018Sub-hertzLaser}. The waveguide samples are tapered to match standard single mode fiber (SMF) at the chip edge, resulting in a coupling loss of 1.5 dB per facet. The measured propagation lossess of the SDS and ADS waveguides are 0.22~dB/cm and 0.15~dB/cm, respectively.

The backward SBS process in the multilayer silicon nitride waveguide structures is diagrammed in Fig \ref{fig:fig1} \textbf{(e)}. Two counter-propagating light waves (a probe at $\omega_{1}$ and a stronger pump at a higher frequency $\omega_{2}$) interfere. This interference generates an acoustic wave through electrostriction. This acoustic wave, with a frequency $\Omega=\omega_{2}-\omega_{1}$, then generates a moving grating via the photoelastic effect. Pump light that scatters of this moving grating undergoes a Doppler shift, which makes it match the probe (Stokes) frequency. This leads to a narrow-band gain at the Stokes frequency. In the multilayer silicon nitride waveguide, acoustic wave confinement occurs between the silicon nitride layers leading to enhanced SBS interaction.  

\subsection*{Enhancement and Inhibition of SBS}\label{sec:Sim}

\begin{figure*}[th]
    \centering
    \includegraphics[width=0.9\textwidth]{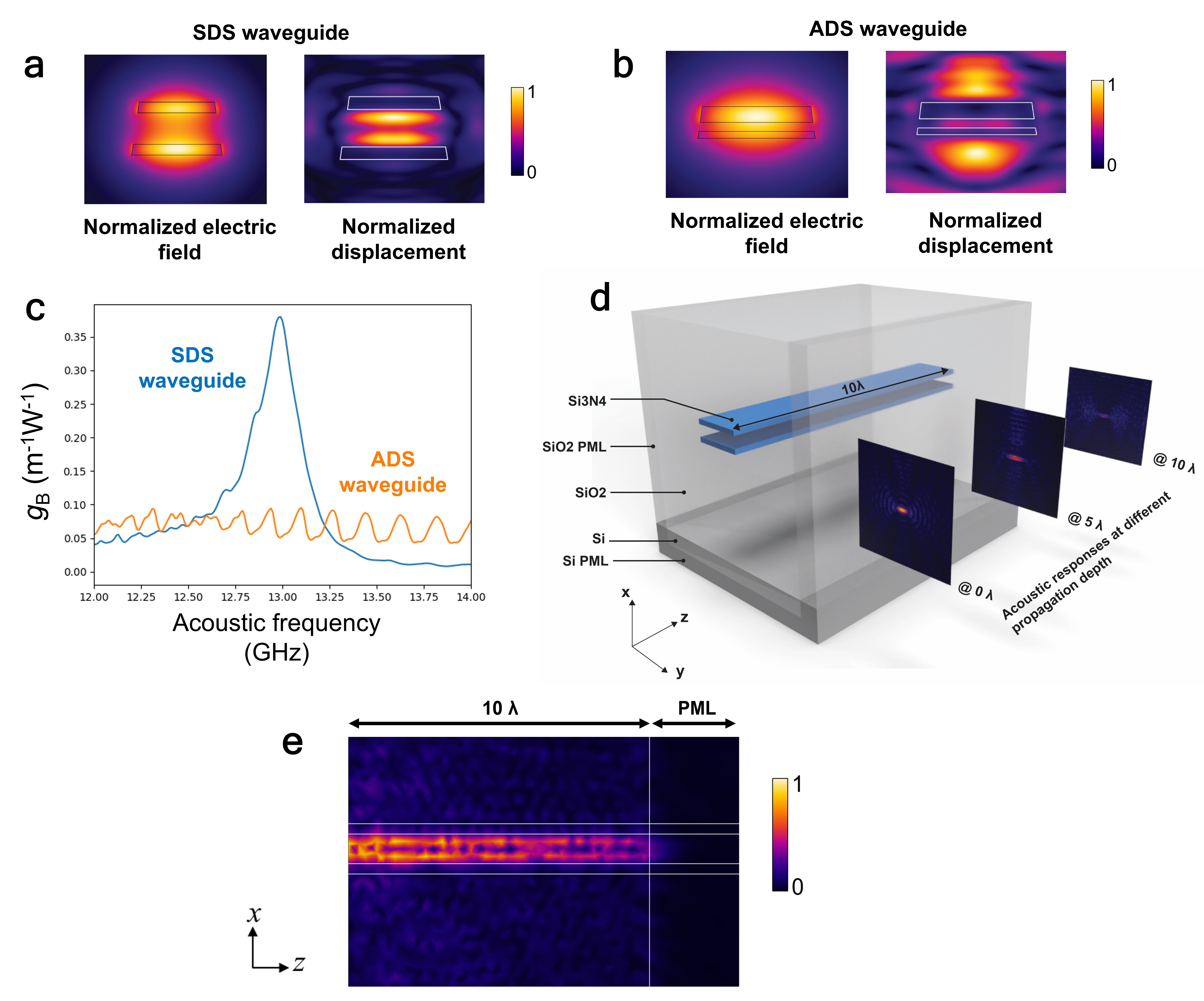}
    \caption{\textbf{Optical and acoustic simulation results of the ADS and SDS waveguides.} \textbf{(a-b)} FEM simulation of the  optical and acoustic mode of the SDS and ADS waveguides with a width $w$ of  $1.2$ µm. Both waveguides have an optical effective index of 1.535. The SDS waveguide shows acoustic confinement in the silicon oxide layer between the silicon nitride layers and enhanced overlap between the optical and the acoustic modes. In contrast, the ADS waveguide shows lack of acoustic confinement and thus reduced opto-acoustic overlap. \textbf{(c)} The calculated SBS gain coefficient of the SDS and ADS waveguides with a width $w=1.2$ µm. The ADS waveguide shows inhibited SBS, with gain coefficient below 0.1 m$^{-1}$W$^{-1}$. The SDS waveguide shows enhanced SBS with a gain coefficient 3-5~times larger than previously demonstrated in silicon nitride.   \textbf{(d)} The 3D acoustic simulation domain used to estimate the acoustic propagation length in the SDS waveguide, with the resulting acoustic field after 0, 5 and 10 wavelengths of propagation (see Supplementary A for details of the simulation). \textbf{(e)} Simulated acoustic field strength as a function of a propagation distance in an SDS waveguide, showing a clear indication of acoustic waveguiding.}
    \label{fig:fig2}
\end{figure*}

To determine the accessible range of SBS performance of the multilayer silicon nitride waveguide structures, we carry out finite element simulations of the optical modes and acoustic modes, along with the SBS gain coefficient, $g_{\rm{B}}$ expressed in m$^{-1}$W$^{-1}$ , which is a function of the overlap between the optical and acoustic modes (see Supplementary~A).   

Fig.~\ref{fig:fig2}~\textbf{(a)} and \textbf{(b)} depict the simulated optical and acoustic modes of $1.2$~µm-wide SDS and ADS waveguides, respectively. When comparing the acoustic responses of the two multilayer waveguides, we observe very distinct behaviours. For the SDS waveguide, strong acoustic response at the frequency of 12.99 GHz was observed in between the 500~nm-separated silicon nitride layers. Such a confinement of the acoustic wave significantly increases the overlap of this acoustic mode with the fundamental TE optical mode shown in  Fig.~\ref{fig:fig2}~\textbf{(a)}. In contrast, the ADS waveguide with only 100~nm separation between the silicon nitride layers, shows no acoustic confinement. At the frequency of  12.99~GHz, the acoustic response of the ADS waveguide resides mostly outside the region of the optical waveguide, thereby significantly reducing the overlap between this acoustic mode with the TE fundamental optical mode. 

As a result of the difference in acoustic waveguiding behaviours, the simulated SBS gain coefficients of the SDS and the ADS waveguides show significant differences, as depicted in Fig.~\ref{fig:fig2}~\textbf{(c)}. The SDS waveguide shows a clear peak at 12.99~GHz with $g_{\rm{B}}=0.38$~ m$^{-1}$W$^{-1}$. This calculated value is 3-5~times larger than previously reported in thin \cite{Gundavarapu2018Sub-hertzLaser} and thick \cite{Gyger2020ObservationWaveguides} silicon nitride waveguides, indicating the potential of enhancing SBS in multilayer waveguides. In contrast, the simulated gain spectra of the ADS waveguide shows no clear SBS peak, with the calculated $g_{\rm{B}}$ stays below 0.1~ m$^{-1}$W$^{-1}$ over a broad frequency range of 12-14 GHz. We believe that the absence of acoustic guiding in the ADS structure with relatively thin separation between the silicon nitride layers (approximately 5 times smaller than the acoustic wavelength) leads to inhibition of the SBS in this structure. Remarkably, this stark difference in SBS response occurs in two waveguides structures with very similar optical performance. Both the SDS and ADS waveguides has very similar effective index, bend radius, and propagation loss. This is highlighting the importance of acoustic waveguiding for SBS actuation and points to advanced control of SBS through tailoring of  waveguide cross-section. 

In order to verify the acoustic waveguiding properties of the SDS waveguide, we developed a 3D acoustic simulation model, as depicted in Fig.~\ref{fig:fig2}~\textbf{(d)}. We applied the acoustic response obtained from the 2D model (see Supplementary Note A for details) to the \textit{xy}-plane at $z=0$, and propagated the acoustic response over 10 wavelengths in the $z$ direction, after which a perfectly matched layer (PML) is applied to prevent any acoustic backscattering. The plots of acoustic field at various positions along the propagation direction are also depicted in Fig.~\ref{fig:fig2}~\textbf{(d)}. A more comprehensive picture of the acoustic waveguiding is shown in Fig.~\ref{fig:fig2}~\textbf{(e)} where the \textit{xz}-plane cut of the SDS waveguide is depicted. The acoustic field stays in between the silicon nitride layers beyond more than 5 µm of propagation (10~× the acoustic wavelength), clearly showing acoustic waveguiding.

\subsection*{Experimental Results}\label{sec:width}

\begin{figure*}[p]
    \centering
    \includegraphics[width=0.85\textwidth]{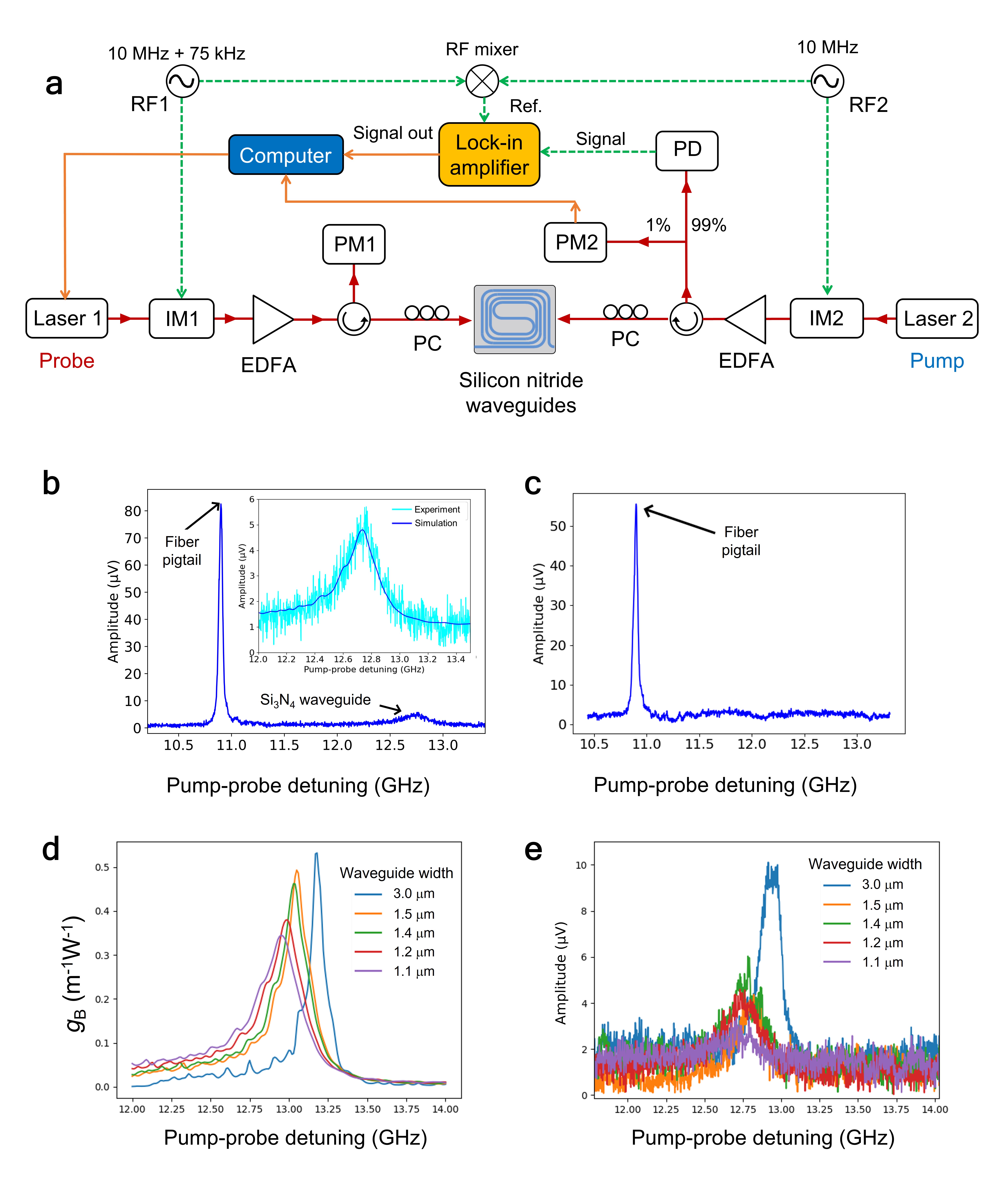}
    \caption{\textbf{Characterisation of SBS in multilayer silicon nitride waveguides.} \textbf{(a)} Schematic of the setup used for the SBS characterization. IM: Intensity modulator, EDFA: erbium-doped fiber amplifier, PM: optical power meter, PC: fiber polarization controller, PD: photodetector, RF: radiofrequency signal generator. See Methods for details of the experiment. \textbf{(b)} Measurement of the SBS gain spectra  from a 50-cm SDS spiral waveguide with $1.2$ µm width. The highest peak at 10.9~GHz is the SBS response of the 3.5 m-long fiber pigtail in the setup with a gain coefficient of 0.14 m$^{-1}$W$^{-1}$. The lower peak around 12.75~GHz is the chip SBS response. Inset: zoomed in view of the experimental result, with the simulation result overlaid. \textbf{(c)} Measured SBS gain spectra from a 50~cm-long, $1.2$~µm~wide ADS waveguide. Only the fiber SBS response at 10.9 GHz is visible here, indicating inhibition of SBS in the ADS waveguide.  \textbf{(d)} Simulation results of the SBS gain spectra in SDS waveguides for the varying waveguide widths represented in the available samples. Highest gain coefficient of 0.53 m$^{-1}$W$^{-1}$ was predicted for the $3$~µm-wide waveguide. \textbf{(e)} Measured SBS gain spectra for SDS waveguides with varying waveguide widths, showing very good agreement with the trend predicted from simulations. The waveguide loss was  ranging from 0.19--0.23 dB/cm. The highest gain of 0.40 m$^{-1}$W$^{-1}$ was measured from the $3$~µm-wide waveguide.}
    \label{fig:fig3}
\end{figure*}

\begin{table*}[t]
\caption{Simulated and measured Brillouin properties of SDS silicon nitride waveguides.}
\label{tab:res}
\begin{tabular}{c|c|c|c|c|c|c|c|c}
    \textbf{Waveguide} & \textbf{Propagation} & \textbf{Effective} & \multicolumn{2}{c|}{\textbf{SBS gain coefficient}} & \multicolumn{2}{c|}{\textbf{Brillouin shift}} & \multicolumn{2}{c}{\textbf{Brillouin linewidth}}\\
    \textbf{width} & \textbf{loss} & \textbf{length} & \textbf{Simulated} & \textbf{Measured} & \textbf{Simulated} & \textbf{Measured} & \textbf{Simulated} & \textbf{Measured}\\
    (µm) & (dB/cm) & (cm) & (m$^{-1}$W$^{-1}$) & (m$^{-1}$W$^{-1}$) & (GHz) & (GHz) & (MHz) & (MHz)\\
    \hline
    1.1 & 0.228 & 17.67 & 0.34 & 0.20 & 12.95 & 12.68 & 380 & 490\\
    1.2 & 0.223 & 17.98 & 0.38 & 0.24 & 12.99 & 12.74 & 330 & 290\\
    1.4 & 0.206 & 19.11 & 0.46 & 0.25 & 13.03 & 12.76 & 260 & 250\\
    1.5 & 0.230 & 17.55 & 0.50 & 0.25 & 13.05 & 12.81 & 230 & 180\\
    3.0 & 0.195 & 19.91 & 0.53 & 0.40 & 13.18 & 12.93 & 130 & 130
\end{tabular}
\end{table*}

We devised an experimental apparatus diagrammed in Fig.~\ref{fig:fig3}~\textbf{(a)} to measure SBS in our silicon nitride samples. The signal magnitudes predicted from our simulations warranted high sensitivity from the experiments. For this reason, we resorted to the dual intensity modulation (IM) technique, which has previously been used in Brillouin spectroscopy \cite{Grubbs1994HighSpectrometer}, and later on adapted for the detection of SBS in thick silicon nitride waveguides \cite{Gyger2020ObservationWaveguides}. This technique relies on the modulation of both the pump and probe, at slightly different frequencies (10~MHz and 10~MHz + 75~kHz, respectively in our case). Any signal that results from the interaction of the two sources such as SBS will therefore appear at the difference frequency (75~kHz), whereas all signals coming from either the pump or the probe will appear at their respective modulation frequencies. This allows for the use of a lock-in amplifier to filter the signal in the electrical domain, rather than to use optical filtering techniques (see Supplementary Note B for the details of the experiments). 

Fig.~\ref{fig:fig3}~\textbf{(b)} shows the SBS gain response measured from the SDS sample. We observed a signal at 10.9~GHz generated by SBS in 3.5~m-long fiber pigtails in the measurement setup. We have estimated that the SBS gain coefficient of this pigtails to be 0.14 m$^{-1}$W$^{-1}$ \cite{Kobyakov2010StimulatedFibers}. At a distinct frequency of 12.75~GHz we observed the SBS peak from the SDS spiral. We further determine the SBS gain coefficient of the SDS waveguide from the lock in amplifier signal \cite{Gyger2020ObservationWaveguides}. By comparing the peak voltages from the fiber pigtail and the waveguide (see Supplementary Note B for details), we estimated the SBS gain coefficient of the $1.2$~µm-wide SDS waveguide to be 0.24 m$^{-1}$W$^{-1}$, a factor of 2.5 higher compared to the previously measured in silicon nitride waveguides \cite{Gundavarapu2018Sub-hertzLaser, Gyger2020ObservationWaveguides}. 

The Brillouin shift observed in the SDS sample of 12.75~GHz can easily be distinguished from the Brillouin shift of the standard single mode  fiber (SMF) pigtails (10.9~GHz) unlike in the case of thin silicon nitride waveguides \cite{Gundavarapu2018Sub-hertzLaser}. This leads to ease of sample characterisation, removing the need of splicing different types of fibers, such as high numerical aperture (HNA) fibers as was done in previous work \cite{Gundavarapu2018Sub-hertzLaser}. 

The inset of Fig.~\ref{fig:fig3}~\textbf{(b)} compares the simulation and measurement results of SBS in the SDS waveguide, showing excellent agreement paticularly in the lineshape and linewidth of the gain response. A slightly modified slope on the lower frequency of the gain response indicates presence of multiple acoustic modes that creates an asymmetric gain response. 

We performed a similar dual-IM pump-probe measurement on the $1.2$~µm-wide ADS waveguide sample. As depicted in Fig.~\ref{fig:fig3}~\textbf{(c)}, the only measurable signal over the frequency range of 10.5--13 GHz was SBS from the fiber pigtails and not SBS signature coming from the sample. This experimental result confirms the simulation results that predicted inhibition of SBS in the ADS sample due to lack of acoustic waveguiding.

A pathway to enhanced SBS is through reduction of acoustic leakage. The acoustic waveguiding mechanism in the SDS waveguide is similar to waveguiding in  slab optical waveguide where impact of lateral leakage can be reduced by making the waveguide wider. We study the impact of varying the width of the SDS waveguide to SBS gain coefficient through a series of simulations. The results are depicted on Fig.~\ref{fig:fig3}~\textbf{(d)}. Solely by altering the waveguide width, a relatively straightforward step in the fabrication process, while keeping the layer stack unchanged, one can increase the SBS gain coefficient up to 0.53~m$^{-1}$W$^{-1}$ for a 3.0 µm-wide waveguide. This represents nearly an order of magnitude improvement of gain coefficient when compared to results reported earlier. Moreover, the improvement of acoustic waveguiding in wider waveguides is evident from the significant linewidth narrowing of the SBS gain response.  

We fabricated SDS waveguides with different widths and measured their SBS responses. The results are shown in Fig.~\ref{fig:fig3}~\textbf{(e)}. The experimental results confirmed the trend predicted from simulations, in which increasing the waveguide width leads to enhanced SBS gain coefficient, reduces the SBS linewidth, and increases the SBS frequency shift.   

Table~\ref{tab:res} summarized the simulated and the experimentally measured Brillouin gain properties of the SDS waveguide geometries investigated in this work. We observed excellent agreement between the simulated and meaured values for the SBS frequency shift and Brillouin linewidth. The maximum difference between the measured and simulated values of the Brillouin frequency shift was 270 MHz, which is only 2$\%$ of the SBS shift (13 GHz). The maximum variation of the SBS linewidth frequency was 110 MHz, observed in the narrowest waveguide of 1.1 µm. This accounts for 28 $\%$ of the simulated SBS linewidth of 380 MHz. For the rest of the samples, the discrepancies between the simulated and the measured linewidths were below 50 MHz. We observed larger discrepancies in the measured and simulated SBS gain coefficient. We consistently measured lower gain coefficient in our samples ranging from 25-50 $\%$ lower than what was predicted from simulations. We have to note that similar trends and accuracy were also present in previously reported SBS experiments in chip-scale devices including waveguides in chalcogenides \cite{Sturmberg2019FiniteWaveguides}, silicon \cite{Sturmberg2019FiniteWaveguides}, and silicon nitride \cite{Gundavarapu2018Sub-hertzLaser, Gyger2020ObservationWaveguides}. We suspect the presence of waveguide bends in our samples, accounting for a significant portion of the total waveguide length might play a role in reducing the SBS gain coefficients. 

\subsection*{RF Notch Filter Demonstration}

\begin{figure*}[t]
    \centering
    \includegraphics[width=0.9\textwidth]{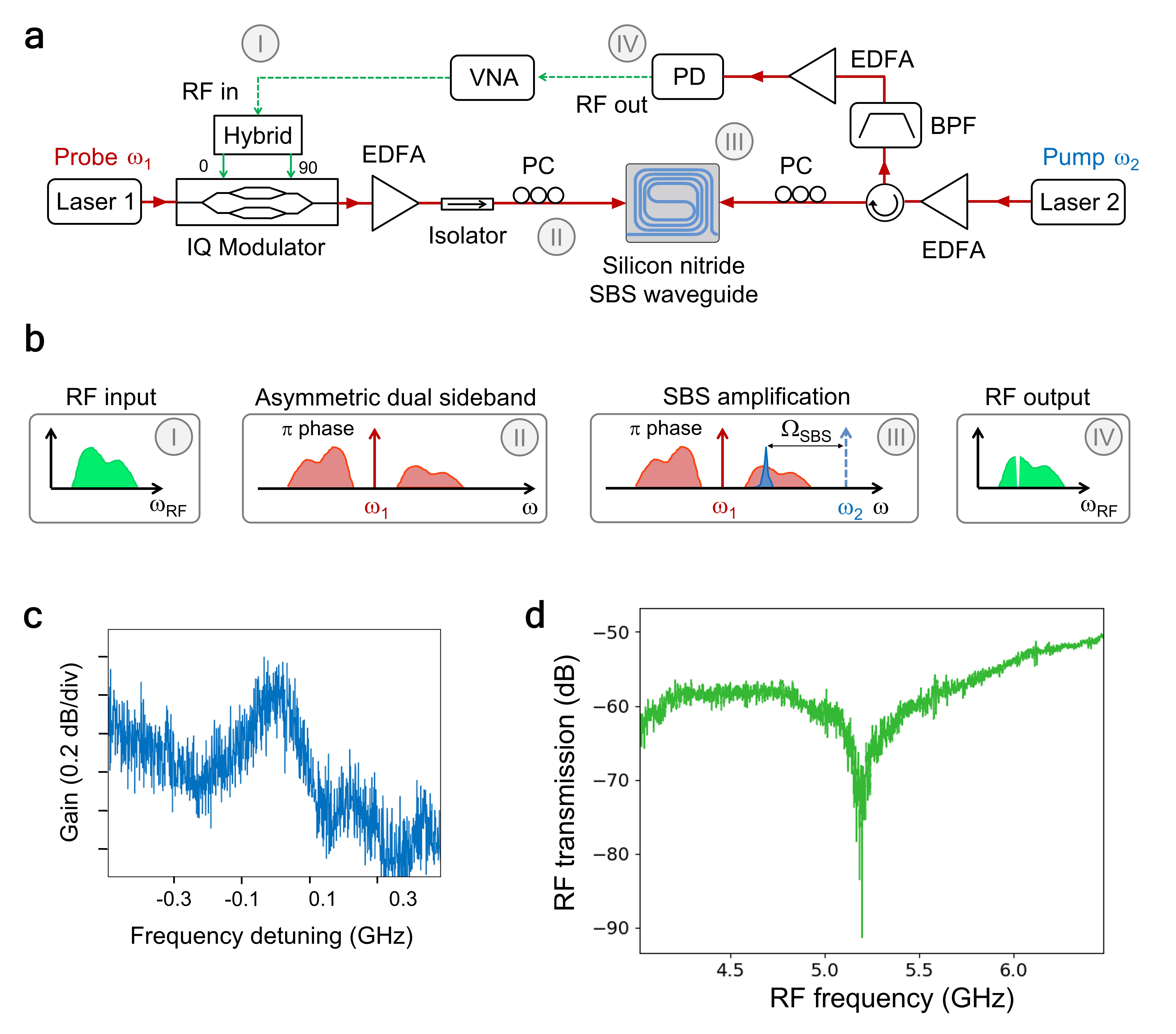}
    \caption{\textbf{Microwave photonic notch filter using SBS in silicon nitride waveguide.} \textbf{(a)} Schematic of the experimental setup to demonstrate the MWP notch filter. The input RF signal generated from a vector network analyzer (VNA) is modulated onto the optical probe using an in-phase quadrature (IQ) modulator. An optical pump is amplified using an erbium-doped fiber amplifier (EDFA) and then injected to $1.4$~µm-wide, 50~cm-long SDS waveguide with 0.25 m$^{-1}$W$^{-1}$ SBS gain coefficient and 0.2 dB/cm propagation loss. An optical bandpass filter (BPF) was used before photodetection (PD) to remove pump back-reflection. Hybrid: $90^{\rm{o}}$ RF hybrid coupler, PC: fiber polarization controller. \textbf{(b)} RF and optical spectra at different points of signal path: (I) Input RF signal. (II) Asymmetric dual sideband phase modulated signal generated from the IQ modulator. The RF sidebands are out of phase and their amplitude ratio is controlled to match the SBS gain magnitude generated from the silicon nitride waveguide.  (III) SBS gain from the silicon nitride waveguide is used equalize the sideband amplitue at the intended RF notch frequency. (IV) At the detector the mixing products between sidebands and the optical carrier leads to a notch filter due to RF cancellation. \textbf{(c)} The measured SBS gain when the IQ modulator is set to single-sideband modulation. \textbf{(d)} The measured high-rejection RF notch filter response which was obtained using only 0.4 dB of on-chip SBS gain. The 3-dB bandwidth of the filter is 400 MHz and the rejection is 30 dB.}
    \label{fig:fig4}
\end{figure*}

One of the key applications of chip-based SBS is microwave photonic signal processing \cite{Marpaung2019IntegratedPhotonics,Gertler2020TunableSilicon,Liu2020IntegratedFilters}. The narrowband feature of an SBS resonance is advantageous for filtering high-density radiofrequency signals. For this reason, we experimentally investigate the feasibility of building an RF notch filter from the SBS gain in a multilayer silicon nitride waveguide. Considering the limited SBS gain from the waveguides, the RF cancellation notch filter \cite{Marpaung2015Low-powerSelectivity, Marpaung2013FrequencyRejection,Marpaung2013Si3N4Rejection}, a scheme that relies on synthesizing destructive RF interference due to careful tailoring of phase and amplitude of RF modulated sidebands, would be an ideal choice. A simplified schematic of the RF photonic notch filter is diagrammed in  Fig.~\ref{fig:fig4}~\textbf{(a)}. A key component in this filter is the in-phase quadrature (IQ) modulator (also known as the dual-parallel Mach-Zehnder modulator) used often synthesizing the RF moudulates sidebands with the correct phase and amplitude relations prior to the narrowband processing using the SBS gain resonance. The details of the RF notch filter experiments can be found in Methods and Supplementary Note B. 

Fig.~\ref{fig:fig4}~\textbf{(b)} shows the working principle of the RF notch filter. The RF input (I) is modulated onto the probe laser using the IQ modulator creating an asymmetric dual sideband modulation, with the sidebands in anti-phase (II). The on-chip SBS interaction with the probe light then amplifies a spectral region of the lower sideband (III) making the sidebands equal in amplitude only at the frequency of the SBS peak gain.  The processed signal was then sent to a photo-diode, resulting in an RF spectrum with a notch response due to the destructive interference between the mixing products of the sidebands and the optical carrier (IV).

The first step in creating the notch filter is to characterize the peak SBS gain exhibited from the waveguide. We achieved this by tuning the IQ modulator to create a single sideband modulation with optical carrier spectrum. The measured SBS gain response obtained from the vector network analyzer (VNA) is depicted in Fig.~\ref{fig:fig4}~\textbf{(c)}, showing a peak gain of approximately 0.4~dB. We then tuned the IQ modulator to synthesize the asymmetric dual sideband modulation with 0.4 dB difference in the two sidebands amplitude. The resulting RF notch filter response is depicted in Fig.~\ref{fig:fig4}~\textbf{(d)}. The peak rejection of the filter was measured to be 30 dB and the filter's 3 dB bandwidth was 400 MHz. This result constitutes the first ever signal processing demonstration of SBS in silicon nitride waveguides and points towards the potential of unlocking unique Brillouin signal processing capabilities in a mature silicon nitride platform.

\section*{Discussion}

We have observed the first signature of SBS in multilayer silicon nitride waveguides. The multilayer waveguides can exhibit enhancement or inhibition of SBS depending on the waveguide cross-section. A waveguide geometry with  enhanced SBS shows acoustic waveguiding in betweeen the silicon nitride layers, thereby increasing the optoacoustic overlap. The geometry with inhibited SBS has a thinner silicon nitride layers separation, preventing acoustic waveguiding. We have developed a simulation framework that can accurately and reliably predict the SBS responses in a various silicon nitride waveguide geometries. We show through simulations and experiments that tailoring the width of the silicon nitride waveguide width can lead to improvement in acoustic waveguiding and enhancement of the SBS gain. We also show that early an order of magnitude gain enhancement compared to previously reported results can be obtained. 

We show that the measured SBS reponse in our silicon nitride waveguides can readily be used for RF photonic signal processing and filtering. Using a mere 0.4 dB of SBS peak gain from the SDS waveguide we demonstrated a n RF notch filter with a 30 dB extinction and a 400 MHz spectral resolution. A drawback from such a low SBS gain is the low RF transmission at the filter passbands, due to partial RF cancellation \cite{Marpaung2015Low-powerSelectivity}. The passband transmission can be significantly improved by combining the SBS resonance with a resonance from an optical ring resonator in an overcoupling state, This has been previously explored using SBS in optical fibers \cite{Liu2016LosslessFilter} or in chalcogenide waveguides \cite{Liu2020IntegratedFilters}. Similar approach can readily be implemented in our silicon nitride platform, with previously demonstrated ring resonator can show  a spectral linewidth as low as 190 MHz \cite{Marpaung2013Si3N4Rejection}. 

\subsection*{SBS Optimization}

\begin{figure}[t]
\centerline{\includegraphics[width=0.48\textwidth]{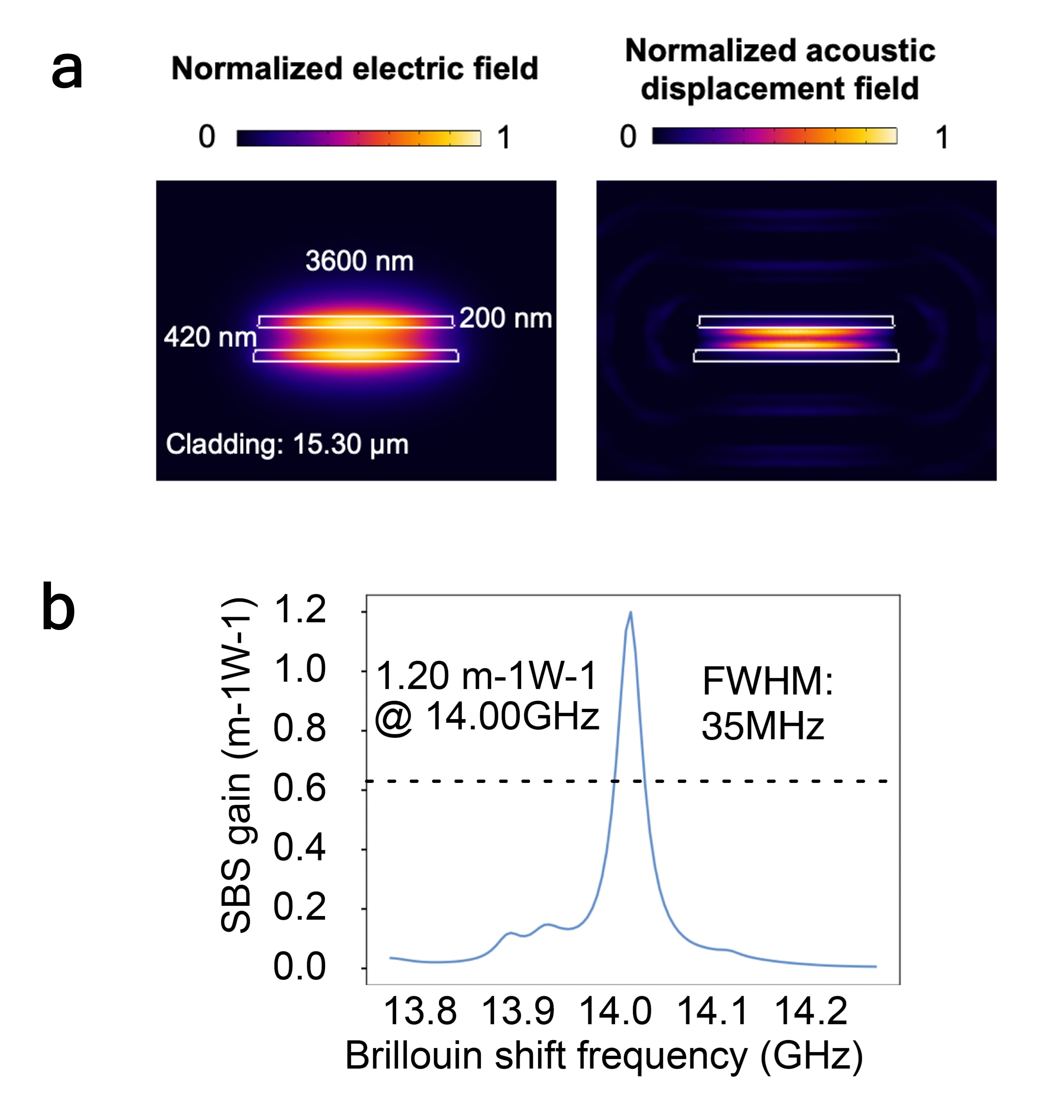}}
\caption{\textbf{Optimization of SBS in SDS waveguide.}  \textbf{(a)} The  optical and acoustic modes of the optimized SDS waveguide geometry obtained from genetic algorithm optimizations. The  waveguide has a width of $3.6$ µm and  silicon nitride layer thickness of 200~nm separated by 420~nm of silicon oxide layer. \textbf{(b)} The calculated SBS gain coefficient of the waveguide of 1.2~m$^{-1}$W$^{-1}$, which is a factor of 12-15 higher than previously reported results. The SBS linewidth of 35 MHz indicates strong acoustic confinement. }
\label{fig:fig5}
\end{figure}

The gain coefficient measured in our waveguides are still relatively low when compared to SBS gain from chalcogenide or silicon waveguides. This stems from the SBS generation mechanism in the multilayer silicon nitride waveguides where the optoacoustic interactions take place primarily in the silicon oxide interlayer. The relatively low refractive index of silicon oxide (n= 1.5) as compared to arsenic trisulfide (n=2.4) or silicon (n=3.45)  has a strong impact on the SBS gain that scales with $n^8$ \cite{Gyger2020ObservationWaveguides}. But when optimized, the SBS actuation in these multilayer waveguides can have a major advantage compared to previous technique because of the maturity and scalability of the underlying waveguide technology. 

We have applied genetic algorithm to optimize the geometry of the muyltilayer silicon nitride waveguide. The details of the optimization procedure can be found in Supplementary Note A. The highest SBS gain obtained from this procedure is of the order of 1.2 m$^{-1}$W$^{-1}$ at  with a SBS frequncy shift of 14~GHz and SBS linewidth of 35~MHz. This represents 15~× gain enhancement compared to the previously reported values \cite{Gyger2020ObservationWaveguides}.  Fig.~\ref{fig:fig5} depicts the normalized electric fields, normalized acoustic displacement fields, and gain profiles of the optimized geometry,  showing strong acoustic confinement.  

When realized, the optimized waveguide geometry can be the basis of unique SBS devices embedded in large-scale and low-loss silicon nitride circuit. The waveguide has a relatively large effective refractive index of 1.615 so they can be used to create bend waveguides with relatively small radii. This makes it possible to build an on-chip SBS laser with short ring length, unlike the SBS lasers realized with thin silicon nitride waveguides \cite{Gundavarapu2018Sub-hertzLaser}. We have estimated that an SBS laser with a free spectral range (FSR) of 14 GHz, that matched the SBS frequency shift in this optimized waveguide, can exhibit a lasing threshold  20~mW when the waveguide propagation loss is of the order of 0.05 dB/cm (see Supplementary Note C for detailed calculations). This power can be provided from an on-chip pump laser realized in a hybrid III-V silicon nitride technology \cite{Klaver2021SelfLaser}, opening the path way to all-integrated narrow linewidth SBS lasers at various wavelength bands. Finally, integration of the optimized SBS waveguides with advanced microwave photonic spectral shaping circuits \cite{Daulay2020On-chipProcessor,Guo2021VersatileShaper} can lead to all-integrated high resolution RF filter with low noise figure and ultra-high high dynamic range \cite{Liu2021IntegratedEnhancement}.  




\section*{Methods}

\subsection*{Silicon Nitride Waveguide Fabrication}

Our waveguides are fabricated using the standard LioniX TriPleX process \cite{Roeloffzen2018Low-LossOverview,Worhoff2015TriPleX:Platform}. First, a 8-µm $\rm SiO_2$ layer is grown from wet thermal oxidation of single-crystal silicon substrate at 1000 \textcelsius. Then, low-pressure chemical vapor deposition (LPCVD) is used for the Si$_3$N$_4$ layers, as well as the intermediate SiO$_2$ layer. Afterwards, the waveguides are patterned using contact lithography, and processed with dry etching. Then, the waveguides are covered with an additional SiO$_2$ layer through LPCVD. Because the differences in coefficient of thermal expansion (CTE) between the deposited layer and the silicon substrate would cause internal stress, the crack-free layer thickness for SiO$_2$ in LPCVD is about 1500~nm. After that, plasma-enhanced chemical vapor deposition (PECVD) is used to increase the SiO$_2$ top cladding thickness to a total of 8~µm.

\subsection*{Determining the SBS Gain Coefficient}

The gain observed in an SBS interaction is determined by:
\begin{equation}\label{eq:SBS}
G = e^{g_{\rm{B}}L_{\rm{eff}}P_{\rm{pump}}},
\end{equation}
where $g_{\rm{B}}$ is the Brillouin gain coefficient in m$^{-1}$W$^{-1}$, $L_{\rm{eff}}$ is the effective length, and $P_{\rm{pump}}$ is the pump power in the waveguide.

The effective length of a waveguide can be calculated using:

\begin{equation}
    L_{\rm{eff}} = \frac{1-e^{-\alpha L}}{\alpha}
\end{equation}

where $\alpha$ is the propagation loss, and $L$ is the waveguide length.

By using \eqref{eq:SBS}, and taking the small signal approximation we can calculate the gain coefficient using:

\begin{equation}
g_{\rm{B}, \rm{SDS}} = \frac{V_{\rm{SDS}}}{V_{\rm{fiber}}}\frac{g_{\rm{B}, \rm{fiber}}L_{\rm{eff, fiber}}P_{\rm{pump,fiber}}}{L_{\rm{eff, SDS}}P_{\rm{pump,SDS}}}.
\end{equation}

Here $V$ denotes the signal voltage measured by the lock-in amplifier, the subscripts \rm{fiber} and \rm{SDS} refer to the properties of the fiber and chip used in this experiment.

\subsection*{SBS Gain Characterisations}

We used a Toptica DFB pro BFY laser at 1550 nm as the probe laser in the experiments. We scanned the laser wavelength using current control, the electric coefficient of this laser is 0.8 GHz/mA. The pump laser is an Avanex A1905LMI. The light is amplified with Amonics AEDFA-33-B (high power) and AEDFA-PA-35 (low noise) amplifiers. We use a Discovery Semiconductor DSC30S photodiode to convert the signals from the optical to the RF domain. 

In the dual modulation experiment, the modulation signals are generated by a Hewlett-Packard 33120A function/arbitrary waveform generator (set to a sine wave with a frequency of 10.075 MHz), and a Wiltron 69147A Synthesized Sweep Generator (set to a fixed frequency of 10 MHz). To prevent crosstalk from entering the modulators, the reference signal is generated by mixing the synchronized output (square wave) of the AWG with the signal generator's 10 MHz reference output. The mixing of these signals is done with a Mini-Circuits ZFM-3H mixer. The lock-in amplifier is an EG\&G Princeton Applied Research model 5510. The probe is modulated using a Thorlabs LN05S-FC intensity modulator, and the pump is modulated with a Covega LN81S-FC intensity modulator.

\subsection*{RF Notch Filter Experiments}
The probe laser is modulated with an RF signal from the VNA (Keysight model P5007A) to the required asymmetric dual sideband modulation with an in-phase quadrature (IQ) modulator (JDSU Dual Parallel Mach-Zehnder Modulator) in combination with a hybrid coupler (Krytar model 3017360K). This signal is then amplified and sent to the chip. The pump is amplified to 2.4~W, and sent to the chip, where it propagates in the opposite direction to the probe.
Because the required modulation scheme is close to regular phase modulation, the observed signal is very weak. A third, low-noise, amplifier is therefore inserted before the photodiode. Pump light that is back-reflected off the facet will result in more light at the photodiode, and also partially deplete the gain of the amplifier. We prevent these issues by using a bandpass filter (EXFO XTM-50-SCL-U) to block the reflected pump light from entering the amplifier.In these experiments, we used the 1.4 µm wide SDS waveguide with gain coefficient of 0.25 m$^{-1}$W$^{-1}$. 

\section*{Author Contributions}
D.M. and R.B. developed the concept and proposed the physical system.  K.Y., R.S., R.B., Y.K., Y.L., and P~.vd.~S. developed and performed numerical simulations. R.B. performed the SBS characterisation and RF filter experiments with input from K.Y., Y.K., O.D., G.L., J.~vd.~H., and L.K.  K.Y. developed the genetic algorithm SBS optimization. Y.K. developed the SBS laser modeling. E.K., M.H., and C.R. developed and fabricated the silicon nitride samples.  D.M., R.B., K.Y., and Y.K. wrote the manuscript with input from Y.L. D.M. led and supervised the entire project.

\section*{Funding Information}
This project was funded by the Nederlandse Organisatie voor Wetenschappelijk Onderzoek (NWO), project numbers 15702 and 740.018.021

\bibliographystyle{IEEEtran}
\bibliography{library}

\newpage
\onecolumngrid
\beginsupplement
\newpage

\section*{Supplementary Information A: Simulation and Optimization}
This supplementary note describes the setup and method we applied for optical and acoustic simulations. The eigenmode model widely applied in the literature is invalid here because of the normalization problem. To get more accurate results, we calculate the excited acoustic mode induced by optical forces instead. Based on our model, we discuss how different parameters of the waveguide would influence SBS gain. We also discuss the pathway to higher SBS gain by optimizing geometry parameters with genetic algorithm. Since the optimized structure can be fabricated without changing processing steps, the results path the way to achieve higher SBS gain in standard SDS waveguide that can be mass produced. 

\subsection*{Simulation Setup and Method}
\begin{figure*}
    \includegraphics{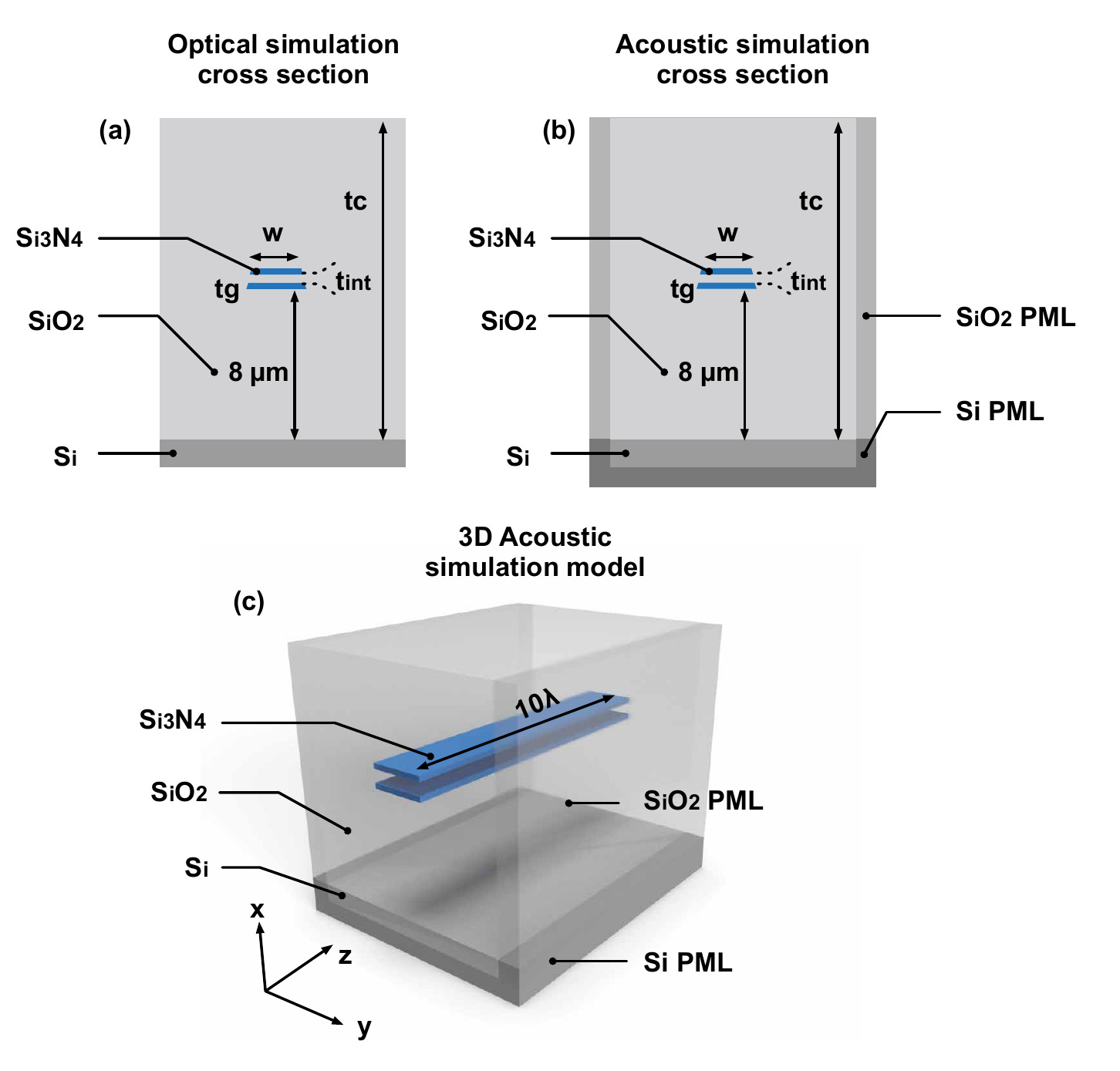}
    \caption{Setup of our simulation model. (a) Optical simulation cross section; (b) Acoustic simulation cross section; (c) 3D acoustic simulation model.}
    \label{fig:model}
\end{figure*}

Fig.~\ref{fig:model} shows the setup of our simulation model. It includes two parts. For the optical simulation, the SDS waveguide is the symmetric double-stripe silicon nitride layer, with stripe thickness of $t_g$, width of $w$, and separation between the two stripes of $t_{int}$. The SDS waveguide is surrounded by the silicon oxide cladding with thickness of $t_c$. Below the cladding is the silicon substrate, which is 8 µm away from the waveguide. We apply COMSOL Multiphysics wave optics solver to simulate the optical modes of the pump and probe. Since the silicon substrate is far away enough from the the waveguide, it has little influence on optical modes. From the optical modes, we can calculate the electrostrictive stress tensor and optical forces induced by electrostriction effect. For the silicon waveguide, radiation pressure also plays an important role in optical forces. However, for the silicon nitride waveguide, since the refractive index between silicon nitride and silicon oxide is much smaller, the influence of radiation pressure is neglected in our model. When optical simulation is finished, we then map the electrostrictive stress tensor and optical forces to the acoustic simulation. 

Since the acoustic mode is not well confined around the waveguide like optical modes, apart from the identical geometry as the optical model, we also need to add perfect match layers (PML) to eliminate reflections at the border in acoustic simulation. One way to calculate the SBS gain is to simulate eigenmodes of the acoustic model, and the total SBS gain is the sum of SBS gain of all individual elastic modes. Since eigenmodes are independent of the acoustic response of the waveguide at certain frequency, this method requires much less computation power and has been widely adopted in the literature\cite{Qiu2013StimulatedGain, Gyger2020ObservationWaveguides}.

The basis of the eigenmode method is that the acoustic response can be decomposed into a finite number of eigenmodes. The condition holds for the silicon waveguide because the acoustic modes are well confined around the waveguide. However, for the silicon nitride waveguide, since the speed of sound in silicon nitride is much higher than in silicon oxide, the acoustic mode is not guided in silicon nitride anymore. Instead, it will radiate into the silicon oxide cladding. As a result, it will be decomposed into a finite number of guided modes and a continuum of radiation modes. Since these radiation modes are not normalizable, the eigenmode method is invalid anymore.

To solve the normalization problem, we calculate the acoustic response of the waveguide at each acoustic frequency point instead. The SBS gain can be calculated from the overlap between acoustic responses ($\tilde{\mathbf{f}}_{\mathrm{n}}$) and optical forces ($\dot{\tilde{\mathbf{u}}}_{\mathrm{n}}$) \cite{Rakich2012GiantLimit}:
\begin{equation}
    G_{B}(\Omega)=2 \cdot\left(\frac{\omega_{s}}{\Omega}\right) \int_{\mathrm{wg}} \operatorname{Re}\left[\tilde{\mathbf{f}}_{\mathrm{n}}(\mathbf{x}, \mathbf{y}) \cdot \dot{\tilde{\mathbf{u}}}_{\mathrm{n}}^{*}(\mathbf{x}, \mathbf{y})\right] d A
\end{equation}
where $\omega_{s}$ is angle frequency of the Stokes and $\Omega$ is the angle frequency of the excited acoustic mode. The cross section of our models matches the real TriPleX SDS waveguide, and material properties applied in our model is listed in \ref{tab:mats}.

We also apply the 3D acoustic simulation model shown in Fig.~\ref{fig:model}~(c) to investigate the propagation of the excited acoustic mode. The acoustic simulation cross section is extruded for 10 wavelengths and followed by PML in $z$ direction. We map the acoustic responses from 2D model as the initial condition and observe the intensity of that field along z direction.

\begin{table}
    \centering
    \caption{Material properties applied in our simulation model.}
    \begin{tabular}{c|c|c|c|c|c}
         &  Refractive Index & Poisson Ratio & Young's Modulus & Density  & p$_{12}$\\
         &  &  & (GPa) & (kg/$\rm m^3$) &\\
         \hline
         Si$_3$N$_4$ \cite{Gyger2020ObservationWaveguides} & 1.98 & 0.23 & 201 & 3020  & 0.047\\
         SiO$_2$ \cite{Smith2016MetamaterialScattering} & 1.45 & 0.19 & 74 & 2240 & 0.27 \\
         Si \cite{Smith2016MetamaterialScattering} & 3.48 & 0.275 & 180 & 2330 & 0.017 \\
    \end{tabular}
    \label{tab:mats}
\end{table}

\subsection*{Influences of different parameters on SBS gain}
\begin{figure*}
    \includegraphics{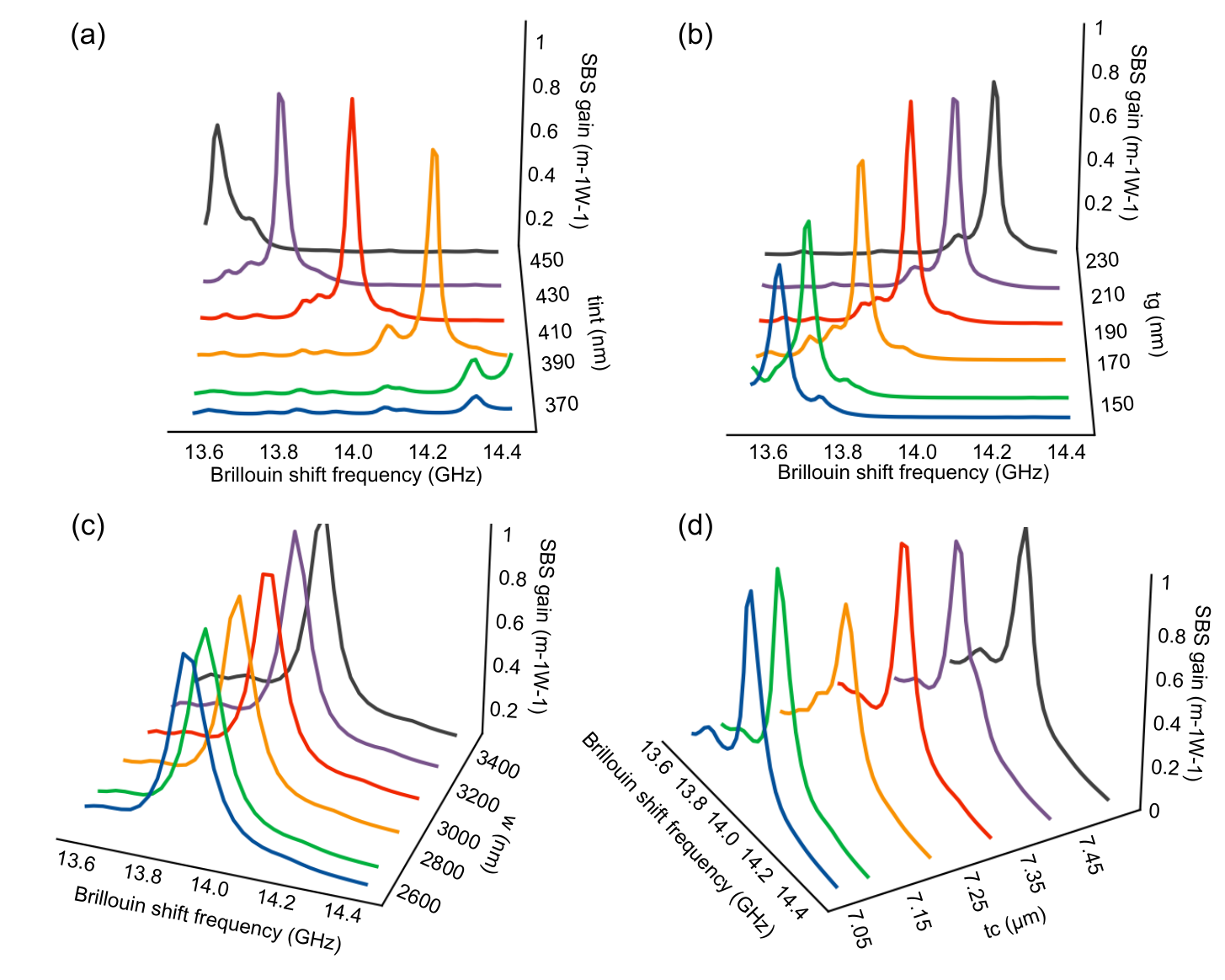}
    \caption{The influences of (a) separation between two stripes; (b) stripe thickness; (c) stripe width on SBS gain profiles; (c) cladding thickness.}
    \label{fig:influence}
\end{figure*}
The SBS gain is related with multiple parameters of the model. To investigate the influences of different parameters, we first find the optimized geometry (with genetic algorithm), then sweep each single parameters to get the corresponding gain profiles. The optimized geometry is when $t_{int} = 420 \rm nm$, $t_{g} = 200 \rm nm$, $t_{c} = 7350 \rm nm$, and $w = 3100 \rm nm$. Fig.~\ref{fig:influence}~(a) - (d) shows influences when we sweep separation $t_{int}$, stripe thickness $t_{g}$, stripe width $w$, and cladding thickness $t_{c}$ separately. First, changing $t_{int}$ would excite different acoustic modes, resulting in different peaks in the gain profile. Second, changing $t_g$ would change the effective index, which would be reflected on different Brillouin shift frequencies. Meanwhile, increasing the width $w$ can improve the confinement of the acoustic wave, however, the effective mode area would increase as well, meaning that the SBS gain cannot be increased by simply increase the width. Finally, the cladding thickness $t_c$ would also influence the SBS gain because the acoustic wave would be reflected at the interface between silicon oxide and air. The SBS gain would be highest when the reflected acoustic wave interferes constructively with the original wave. However, the optimized cladding thickness would change for every different geometry, in a word, we need to optimize these four parameters simultaneously to obtain the optimum geometry that can provide highest SBS gain.

\begin{figure*}
    \includegraphics{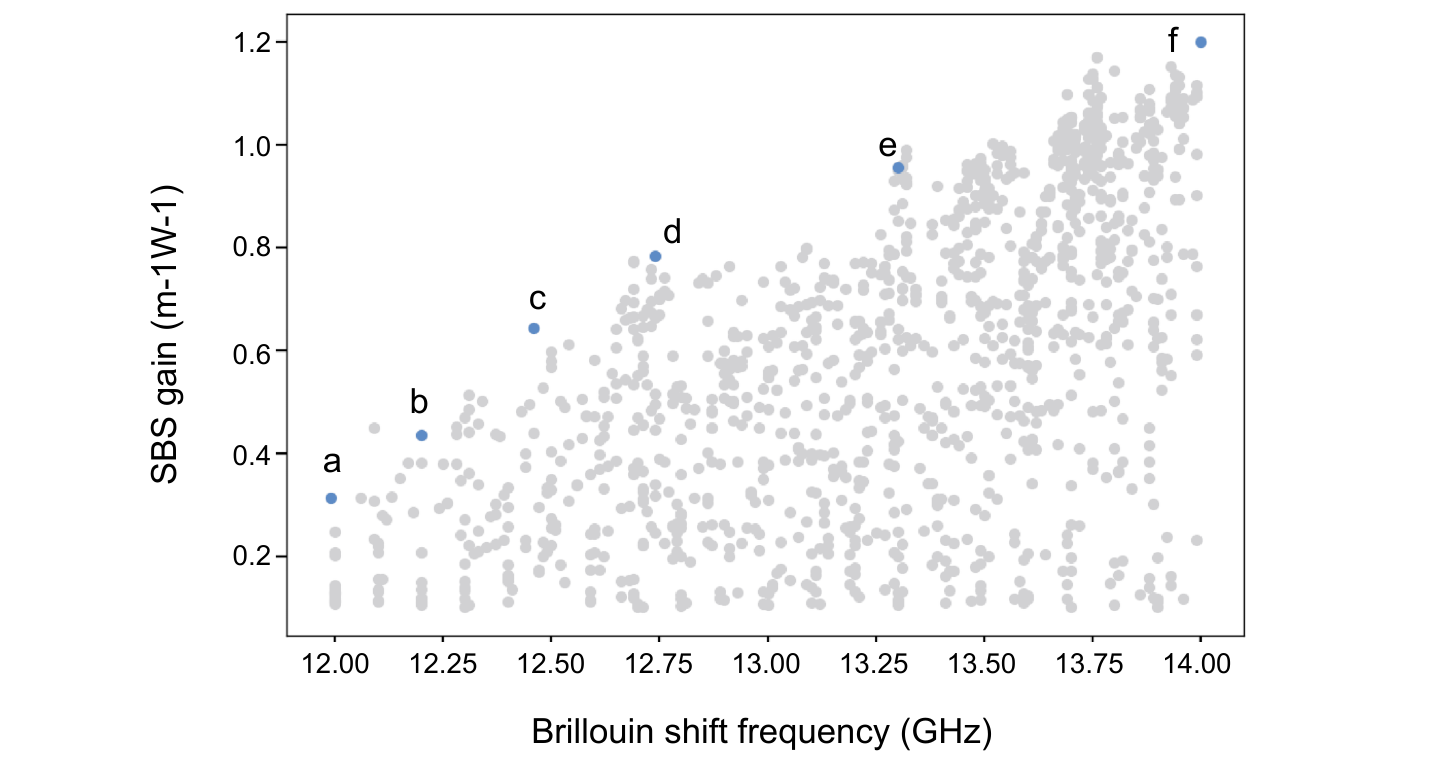}
    \caption{The SBS gain and Brillouin shift frequencies of all simulated geometries  with genetic algorithm. The geometries selected in Fig.~\ref{fig:selectedGeoms} are indicated by blue dots labeled a-f.}
    \label{fig:allselectedGeoms}
\end{figure*}

\subsection*{Optimization with Genetic Algorithm}
We apply the genetic algorithm to optimize the geometry of the waveguide. Table~\ref{tab:ga} shows the gene pool of our algorithm. $t_{g}$ changes from 100 nm to 300 nm with a step of 10 nm, $t_{int}$ changes from 200 nm to 700 nm with a step of 10 nm, $w$ changes from 500 nm to 5000 nm with a step of 100 nm, and $t_{c}$ changes from 15 µm to 17 µm with a step of 20 nm. In total, there are nearly 5 million possibilities. We carried out 14-round evolution in our optimization. The genetic algorithm is carried out as follow. First, we randomly generate 200 candidates from the gene pool. We simulate the SBS gain of each candidate from 12 GHz to 14 GHz. Then, We select 80 elites with the highest peak SBS gain. After that, we generate 80 kids by randomly pairing 80 elites. For each pair, they give birth to two kids. One of them inherits three genes from mother and one gene from father, while the other kids inherits the opposite. During the inheritance, there are 15\% chance that one of the gene would mutate, i.e., the value of that gene would increase or decrease by 20\%. Since the second round, we also introduce 40 new candidates from the gene pool to increase the variety. Finally, both parents and kids and the new candidates would enter the next round evolution. 

\begin{figure*}
    \includegraphics[width=0.9\textwidth]{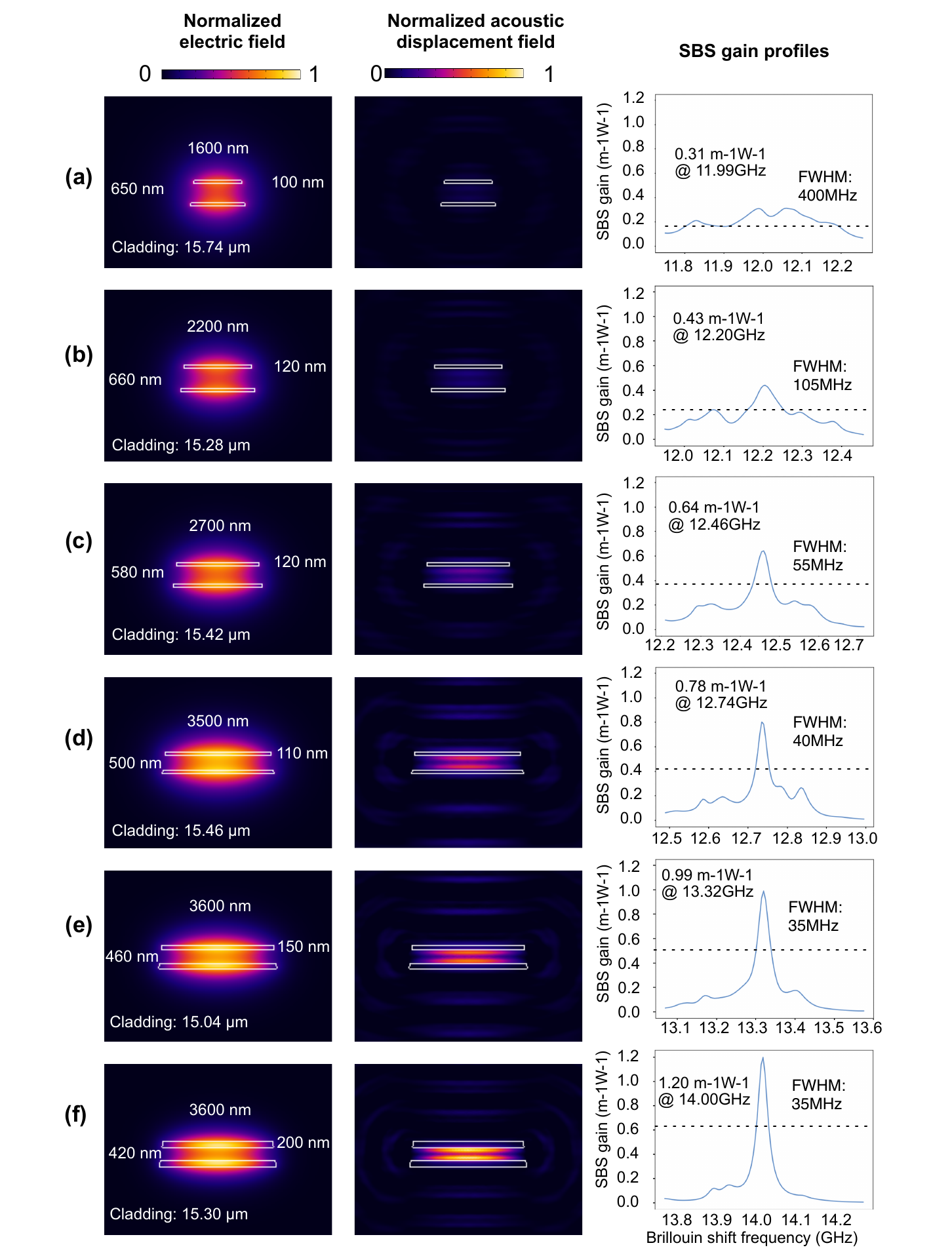}
    \caption{A selection of simulated geometries with SBS gain of 0.31, 0.43, 0.64, 0.78, 0.99, and 1.20 $\rm m^{-1}W^{-1}$. They are indicated in Fig.~\ref{fig:allselectedGeoms} by blue dots labeled a-f. Shown in this figure from left to right are simulated normalized electric field, displacement, and SBS gain profiles. Parameters of each geometry are presented in the normalized electric field.}
    \label{fig:selectedGeoms}
\end{figure*}

Fig.~\ref{fig:allselectedGeoms} shows the SBS gain of all simulated structures with genetic algorithm. The SBS gain tends to increases with the Brillouin shift frequency. The highest gain is around 1.2 m$^{-1}$W$^{-1}$ at 14 GHz. Fig.~\ref{fig:selectedGeoms} shows the normalized electric fields, normalized acoustic displacement fields, and gain profiles of some selected geometries that are labelled (a)--(f) in Fig.~\ref{fig:allselectedGeoms}. As the SBS gain increases, the gain profile also becomes sharper. From the acoustic displacement field, we can also find that the acoustic field is stronger and more confined.

\begin{table}
    \centering
    \caption{Gene pool of our genetic algorithm.}
    \begin{tabular}{c|c|c|c|c}
         \thead{Parameter} & \thead{Range} & \thead{Range step} & \thead{Unit} & \thead{Description}\\
         \hline
        $t_g$ & 100--300 & 10 & nm & Stripe thickness\\
        $t_{int}$ & 200--700 & 10 & nm & Separation between two stripes\\
        $t_c$ & 15-17 & 0.02 & µm & Cladding thickness\\
        $w$ & 500--5000 & 100 & nm & Waveguide width\\
    \end{tabular}
    \label{tab:ga}
\end{table}

\clearpage
\section*{Supplementary Information B: Details of the experiments}

This supplementary note describes parameters used in the experiments and the method used to extract the Brillouin linewidths from the experimental results.

\subsection*{Experimental Parameters}

The main experimental parameters for the dual intensity modulator and notch filter setups can be found in Tables~\ref{tab:SBS_param} and \ref{tab:notch_param} respectively.

\begin{table}[h]
\caption{The experimental parameters of the double intensity modulation setup.}
\label{tab:SBS_param}
\begin{tabular}{c|c|c|c}
    \thead{Parameter} &\thead{Value} & \thead{Unit} &\thead{Description}\\
    \hline
    $P_{probe}$ & 21.3 & dBm & Probe optical power after amplification\\
    $P_{pump}$ & 33.8 & dBm & Pump optical power after amplification\\
    $V_{\pi,probe}$ & 5.5 & V & $V_{\pi}$ of the probe modulator\\
    $V_{\pi,pump}$ & 5.6 & V & $V_{\pi}$ of the pump modulator\\
    $P_{mod,probe}$ & 16.0 & dBm & RF power sent to probe modulator\\
    $P_{mod,pump}$ & 6.0 & dBm & RF power sent to pump modulator\\
    $r_{pd}$ & 0.8 & A/W & Photodiode sensitivity\\
    $\alpha_{c}$ & 2.35 & dB/facet & Coupling loss per facet, including fiber components
\end{tabular}
\end{table}

\begin{table}[h]
\caption{The experimental parameters of the SBS notch filter setup.}
\label{tab:notch_param}
\begin{tabular}{c|c|c|c}
    \thead{Parameter} &\thead{Value} & \thead{Unit} &\thead{Description}\\
    \hline
    $P_{RF}$ & 5.0 & dBm & RF output power of VNA\\
    $P_{probe}$ & 21.3 & dBm & Probe optical power after amplification\\
    $P_{pump}$ & 33.5 & dBm & Pump optical power after amplification\\
    $P_{opt, pd}$ & 13.5 & dBm & Optical power of secondary probe amplifier\\
    $V_{\pi,IQ}$ & 3.5 & V & $V_{\pi}$ of the IQ modulator\\
    $r_{pd}$ & 0.8 & A/W & Photodiode sensitivity\\
    $\alpha_{c}$ & 2.35 & dB/facet & Coupling loss per facet, including fiber components
\end{tabular}
\end{table}

\subsection*{Fitting of the Experimental Results}

\begin{figure*} [b]
    \includegraphics[width=\textwidth]{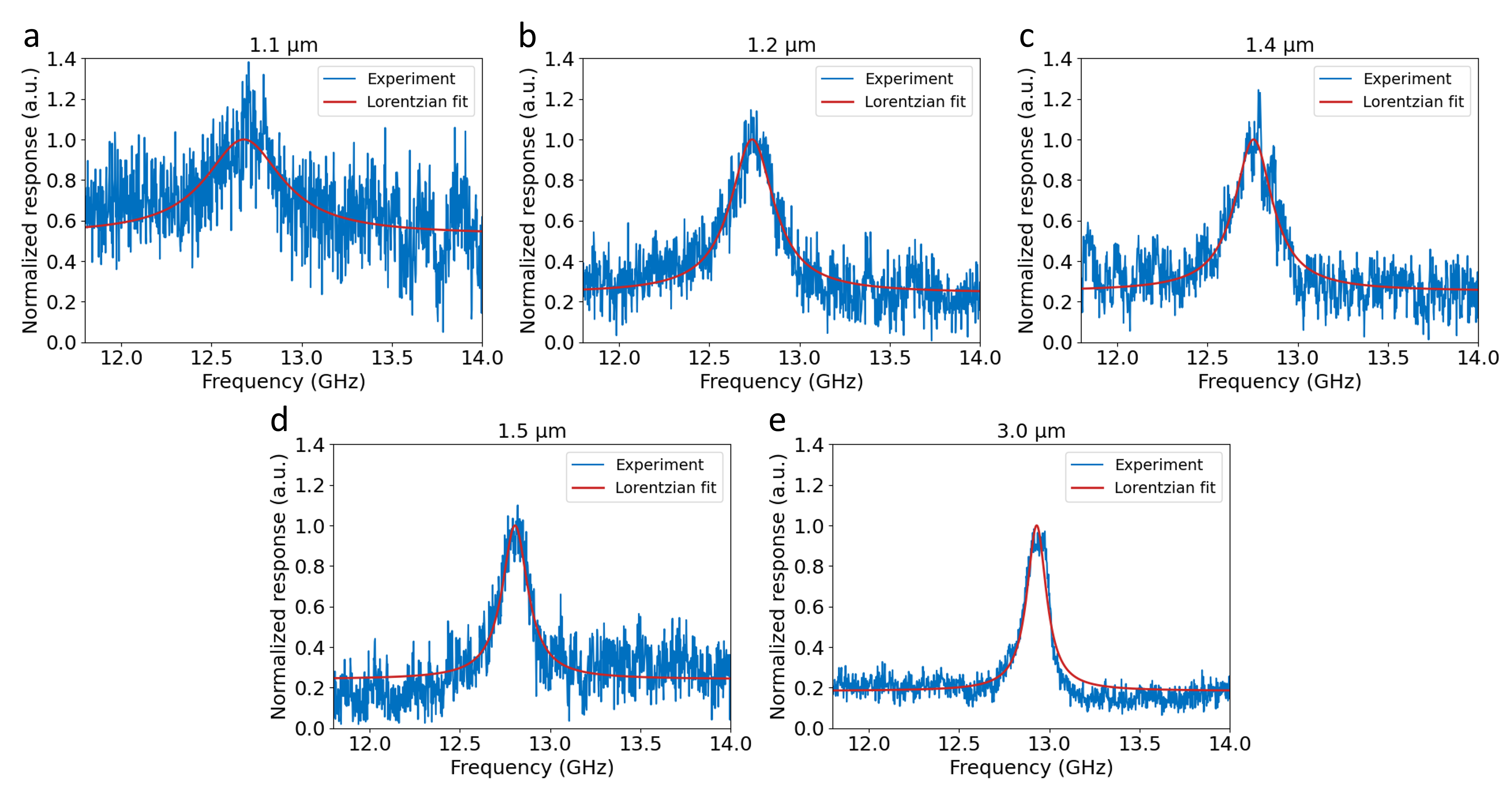}
    \caption{Fitting of the experimental results to extract the Brillouin linewidth values of various waveguide widths, as listed in Table~\ref{tab:res}.}
    \label{fig:exp_fit}
\end{figure*}

To determine the linewidth of our measured Brillouin response we fitted a Lorentzian function to our experimental results. The experimental curves and their fits can be seen in Fig.~\ref{fig:exp_fit}. These fits were made using the Python package \textit{lmfit}, by combining a Lorentzian model with a variable constant, which we applied to take into account the non-zero base-level of our measurements.

\clearpage

\section*{Supplementary Information C: SBS Laser Threshold Calculation}
In this supplementary note we discuss the feasibility of a Brillouin laser based on the waveguide designs described in  Supplementary Note A. To do so we describe the boundary condition for the resonator, the threshold condition and give the loss for different scenarios of propagation loss, from the measured samples and below.

An integrated Brillouin laser consists of a ring resonator pumped resonantly with an external pumping laser whilst on a resonance at the Brillouin downshifted frequency light can build up within the cavity. The restriction of both pump and the SBS shifted light needing to be resonant gives the condition: $\Omega/2\pi = n \cdot FSR_\nu$ with $n$ an integer.  

We can describe the lasing threshold for the first mode \cite{Gundavarapu2018Sub-hertzLaser} and write it as:
\begin{equation}
    P_{th} = \frac{\kappa^{-1}}{g_0 L} \left(\frac{\pi\Delta\nu_L}{FSR_\nu}\right)^2,
\end{equation}
where $\Delta\nu_L$ is the loaded linewidth of the resonator, L the length of the resonator and $\kappa$ the coupling efficiency, not  to be confused with decay rate or coupling strength. The coupling efficiency is given here as the ratio of the energy coupling rate to the total decay rate, i.e., $\kappa = \Delta\nu_{ext}/\Delta\nu_L$. By taking the loaded linewidth to be approximately the sum of the intrinsic linewidth and the coupling rate, i.e., $\Delta\nu_L = \Delta\nu_{ext} + \Delta\nu_0$, the coupling rate can be optimized for minimal lasing threshold as such that the coupling efficiency is $\kappa= 1/3$.

Finally the intrinsic linewidth can be converted to a propagation loss via the mean field approximation, i.e., $ 2\pi \Delta\nu_0 \approx v_g\frac{\alpha}{10\log_{10} (e)}$ with $v_g$ the group velocity and $\alpha$ the propagation loss in dB/cm. Such that, with optimized coupling for minimal threshold, the threshold power can now be expressed as:
\begin{equation}\label{eq:threshold}
    P_{th} = \frac{27}{160\log_{10}(e)}\frac{\alpha^2 L}{g_0}.
\end{equation}

\begin{figure*}
    \includegraphics[width=0.6\textwidth]{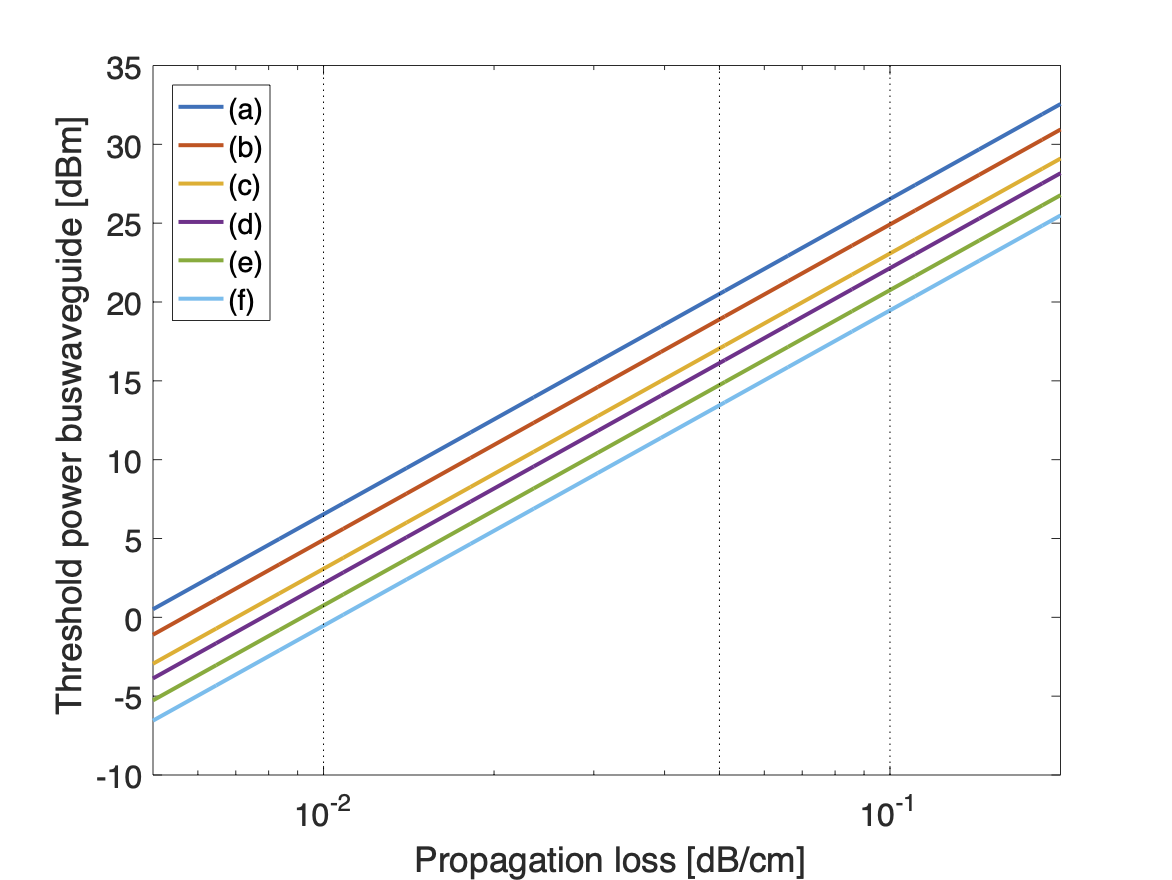}
    \caption{The SBS lasing threshold calculated by equation \ref{eq:threshold} using the parameters from Table~\ref{tab:laser_threshold} for the waveguide geometries a to f given in Fig.~\ref{fig:selectedGeoms} for a loss ranging from 0.2 to 0.005~dB/cm. The dotted lines give the losses 0.1~dB/cm, 0.05~dB/cm, and 0.01~dB/cm, for which the threshold powers are shown in Table~\ref{tab:laser_threshold}.}
    \label{fig:laser_threshold}
\end{figure*}

By applying equation \ref{eq:threshold} on the different gain parameters and frequency shifts of the SBS waveguides a to f selected in supplementary A and shown in Fig.~\ref{fig:selectedGeoms}, with the length such that $\Omega/2\pi = FSR_\nu$, we find a relation as given in Fig.~\ref{fig:laser_threshold} with the details given in Table~\ref{tab:laser_threshold}. The group indices shown in Table~\ref{tab:laser_threshold} have been calculated for the different waveguides via the method given in \cite{Afshar2009ANonlinearity} and are used to convert the target FSR to the length of the resonator by $FSR = \frac{c}{n_g L}$.

\begin{table} [h]
\caption{Laser threshold calculated for waveguide geometries a to f as shown in Fig.~\ref{fig:selectedGeoms} for for a Brillouin laser with a free spectral range matching the Brillouin shift.}
\label{tab:laser_threshold}
\begin{tabular}{c|c|c|c|c|c|c|c}
    \thead{Waveguide} & \thead{Group index} & \thead{FSR} & \thead{Length} & \thead{Brillouin Gain} & \thead{$P_{th}$ for $\alpha = 0.1$ dB/cm} & \thead{$P_{th}$ for $\alpha = 0.05$ dB/cm} & \thead{$P_{th}$ for $\alpha = 0.01$ dB/cm} \\
     geometry &  $n_g$ & (GHz) & (mm) & (m$^{-1}$W$^{-1}$) & (dBm) & (dBm) & (dBm)\\
    \hline
    a & 1.603 & 11.99 & 15.60 & 0.31 & 26.5 & 20.5 & 6.5\\
    b & 1.648 & 12.20 & 14.92 & 0.43 & 24.9 & 18.9 & 4.9\\
    c & 1.656 & 12.46 & 14.53 & 0.64 & 23.1 & 17.1 & 3.1\\
    d & 1.646 & 12.74 & 14.30 & 0.78 & 22.1 & 16.1 & 2.1\\
    e & 1.723 & 13.22 & 13.16 & 0.99 & 20.8 & 14.7 & 0.8\\
    f & 1.804 & 14.00 & 11.87 & 1.20 & 19.5 & 13.4 & -0.5\\

\end{tabular}
\end{table}
\clearpage

\end{document}